\documentclass[aps,prl,nofootinbib,superscriptaddress,twocolumn]{revtex4}
\usepackage{graphicx}
\newcommand{\Dy}{Dy$_{3}$}
\newcommand{\pDy}{Dy$^{3+}$}
\newcommand{\Dyfull}{[Dy$_{3}$($\mu_{3}$-OH)$_{2}$L$_{3}$Cl(H$_{2}$O)$_{5}$]Cl$_{3}$}
\newcommand{\msr}{$\mu$SR}

\begin{document}

\title{Probing the magnetic ground state of the molecular Dysprosium
triangle}

\author{Z.~Salman}
\email{zaher.salman@psi.ch}
\affiliation{Clarendon Laboratory, Department of Physics, Oxford
  University, Parks Road, Oxford OX1 3PU, UK}
\affiliation{Laboratory for Muon Spin Spectroscopy, Paul Scherrer Institut, CH-5232 Villigen PSI, Switzerland}
\author{S. R. Giblin}
\affiliation{ISIS Facility, Rutherford Appleton Laboratory, Chilton,
  Oxfordshire, OX11 0QX, UK}
\author{Y. Lan}
\author{A. K. Powell}
\affiliation{Institut f\"ur Anorganische Chemie, Universit\"at Karlsruhe (TH),
Engesserstr. 15, 76131 Karlsruhe, Germany}
\author{R.~Scheuermann}
\affiliation{Laboratory for Muon Spin Spectroscopy, Paul Scherrer Institut, CH-5232 Villigen PSI, Switzerland}
\author{R.~Tingle}
\affiliation{ISIS Facility, Rutherford Appleton Laboratory, Chilton,
  Oxfordshire, OX11 0QX, UK}
\author{R.~Sessoli}
\affiliation{Dipartimento di Chimica, Universit\'a di Firenze \& INSTM, via della
Lastruccia 3, 50019 Sesto Fiorentino, Italy}

\begin{abstract}
  We present zero field muon spin lattice relaxation measurements of a
  Dysprosium triangle molecular magnet. The local magnetic fields
  sensed by the implanted muons indicate the coexistence of static and
  dynamic internal magnetic fields below $T^* \sim 35$ K. Bulk
  magnetization and heat capacity measurements show no indication of
  magnetic ordering below this temperature. We attribute the static
  fields to the slow relaxation of the magnetization in the ground
  state of \Dy. The fluctuation time of the dynamic part of the field
  is estimated to be $\sim 0.55$ $\mu$s at low temperatures.
\end{abstract}

\maketitle

\section{Introduction}
\begin{figure}[h]
\includegraphics[width=0.6\columnwidth]{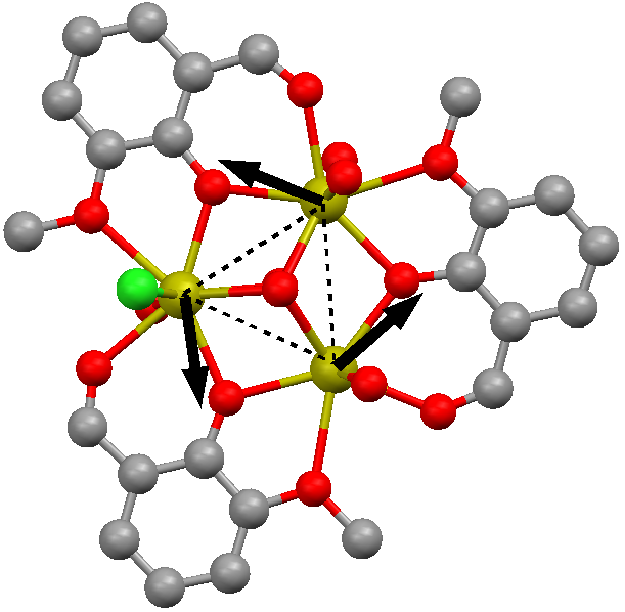}
\caption{(color online) The magnetic core of the \Dy\ molecules. The
  yellow, red, green and gray balls represent the Dy, O, Cl, and C
  atoms, respectively.\label{Dy3Core}}
\end{figure}

Recent development in the field of molecular magnetism has produced
idealized model systems that allow the investigation of quantum
phenomena in nanomagnets \cite{GatteschiMNN}. Examples include slow
relaxation in high spin molecules \cite{Sessoli93N}, quantum tunneling
of the magnetization \cite{Friedman96PRL,Thomas96N}, topological
quantum phase interference \cite{Wernsdorfer99S,Wernsdorfer00EL}, and
quantum coherence \cite{delBarco04PRL,Morello06PRL,Bertaina08N}.
Within these systems, antiferromagnetic clusters, rings or triangles
\cite{Ardavan07PRL,Bertaina08N}, have been considered for possible use
in quantum computation, due to longer decoherence time as a result of
weaker intercluster dipolar interactions \cite{Morello06PRL}. Here we
investigate a recently synthesized trinuclear cluster \Dyfull\ (where
L is the anion of ortho-vanillin) \cite{Tang06ACIE}, hereafter
abbreviated as \Dy\ (see Fig. \ref{Dy3Core}).
This compound exhibits two important properties; a system with a
nonmagnetic ground state as well as slow relaxation of the
magnetization which is typical of some high spin clusters
\cite{Luzon08PRL}. Magnetization measurements have been explained
using a model of three \pDy\ moments arranged on a triangle, each with
a large Ising anisotropy axis in the plane rotated by 120$^o$ relative
to its neighbors (see Fig. \ref{Dy3Core}). A small antiferromagnetic
interaction ($\sim 0.2$ K) between neighboring \pDy\ results in a
nonmagnetic, $S=0$, ground state characterized by a vortex
spin-chirality.

In this paper we present a zero field (ZF) muon spin relaxation (\msr)
study of \Dy.  In this technique polarized muons are used as local
spin probes of the microscopic magnetic structure of the individual
\Dy\ molecules. We find that below $T^* \sim 35$ K the local magnetic
field experienced by the muons has two components, dynamic and static.
The static component is direct evidence of the slow relaxation of the
magnetization in the ground state of \Dy, while the dynamic component
is used to measure the fluctuations time of the \Dy\ ground spin state
directly. This is possible due to the unique local perspective of the
muon.

\section{Experimental}
The \msr\ experiments were performed on the DOLLY spectrometer at the
Paul Scherrer Institute, Switzerland. In these experiments $100 \%$
polarized (along the beam direction, $z$) positive muons are implanted
in the sample. Each implanted muon decays (lifetime $\tau_{\mu}=2.2$
$\mu$sec) emitting a positron preferentially in the direction of its
polarization at the time of decay. Using appropriately positioned
detectors, one measures the asymmetry of the muon beta decay along $z$
as a function of time $A(t)$, which is proportional to the time
evolution of the muon spin polarization. $A(t)$ depends on the
distribution of internal magnetic fields and their temporal
fluctuations. Further details on the \msr\ technique may be found in
Ref.~\cite{Kilcoyne98}.

The composition and structure of \Dy\ sample was confirmed using X-ray
diffraction and magnetization measurements. The powder sample was
placed in a $^4$He gas flow cryostat to measure the muon spin
relaxation in the temperature range between 1.6 and 300 K. Heat
capacity measurements at ZF and in the temperature range between 2 and
100 K were performed on the same sample using a Quantum Design
Physical Properties Measurement System. Additionally, magnetization
measurements in the same temperature range were performed using a
Quantum Design Magnetic Properties Measurement System.

\begin{figure}[h]
\includegraphics[width=0.9\columnwidth]{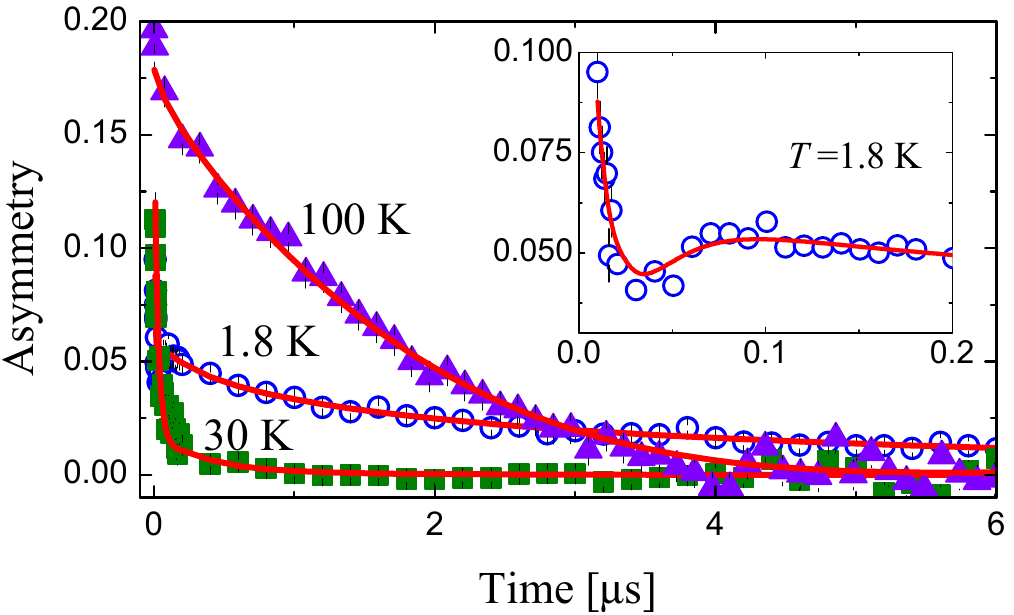}
\caption{(color online) The asymmetry as a function of time at three
  different temperatures.  The inset shows the asymmetry at very short
  times, where the dip is clear evidence for a static local field
  experienced by the muons. The solid lines are fits to
  Eq.~(\ref{AsyLKT}). \label{Asy}}
\end{figure}

Example muon spin relaxation curves measured in ZF are presented in
Fig.~\ref{Asy}. Note that at low temperatures, the asymmetry exhibits
a dip at early times (inset of Fig.~\ref{Asy}), followed by a recovery
and then relaxation at longer times. In contrast, at high temperature
the asymmetry relaxes almost exponentially from its initial value to
zero. The low temperature relaxation curves are evidence that the
internal magnetic field experienced by the implanted muons in \Dy\
contains two contributions; a static (time independent) component and
a fluctuating component. As we discuss below, the static component is
direct evidence of the slow relaxation of the individual \Dy\ (or
\pDy) magnetic moments.
Generally, when muons experience a distribution of static magnetic
fields $\rho \left({\bf B}-{\bf B}_s ,\Delta \right)$, where ${\bf
  B}_s$ is the average static field and $\Delta$ is the root mean
square of the field distribution, then the asymmetry follows a static
Kubo-Toyabe function
\begin{eqnarray} \label{KT}
  A_{\rm KT}(t)&=&A_0\int \rho \left({\bf B}-{\bf B}_s,\Delta \right) G_z(t) d^3B\nonumber \\
  G_z(t)&=& \cos^2 \theta + \sin^2 \theta \cos (\gamma B t).
\end{eqnarray}
Here $\gamma=135.5$ MHz/T is the muon gyromagnetic ratio, $\theta$ is
the angle between the initial muon spin and the local static magnetic
field ${\bf B}$ which is averaged over a powder sample. For example,
if ${\bf B}_s=0$ then the asymmetry is at its maximum value at $t=0$,
it exhibits a dip at $t \sim 1/\Delta$ and recovers to $\sim 1/3$ its
initial value at long times. Depending on the form of $\rho$, e.g.
Gaussian or Lorentzian, the relaxation follows a Gaussian Kubo-Toyabe
(GKT), $A_{\rm GKT}$, or a Lorentzian Kubo-Toyabe (LKT), $A_{\rm
  LKT}$, respectively. However, if in addition to the static field
component a small fluctuating field $B_d(t)$ is present, then the
$1/3$ tail continues to relax to zero
\cite{Hayano79PRB,Uemura85PRB,Uemura98}. In this case the relaxation
can be described by a phenomenological function: a LKT or GKT
multiplied by a suitable dynamic relaxation function
\cite{Fudamoto02PRB}.

In our case, the asymmetry measured in \Dy\ at all temperatures was
found to fit best to LKT multiplied by a square root exponential
relaxation,
\begin{equation} \label{AsyLKT}
A(t)=A_{\rm LKT}(t) e^{- \sqrt{\lambda t}}.
\end{equation}
where $\lambda$ is the relaxation rate, containing information
regarding the dynamics of the local field. In particular, at low
temperatures where $\Delta > \lambda$, $A(t)$ is almost identical to
the well known dynamic LKT function and hence $\lambda=2/3\tau$
\cite{Hayano79PRB,Uemura85PRB}, where $\tau$ is the fluctuation time
of the local magnetic field \cite{footnote}. The square root
exponential relaxation reflects the averaging of the relaxation
behavior of muons stopping in multiple inequivalent sites
\cite{Uemura85PRB,Lascialfari98PRL,Salman02PRB,Blundell03P,Branzoli09PRB}.

\begin{figure}[h]
\includegraphics[width=0.9\columnwidth]{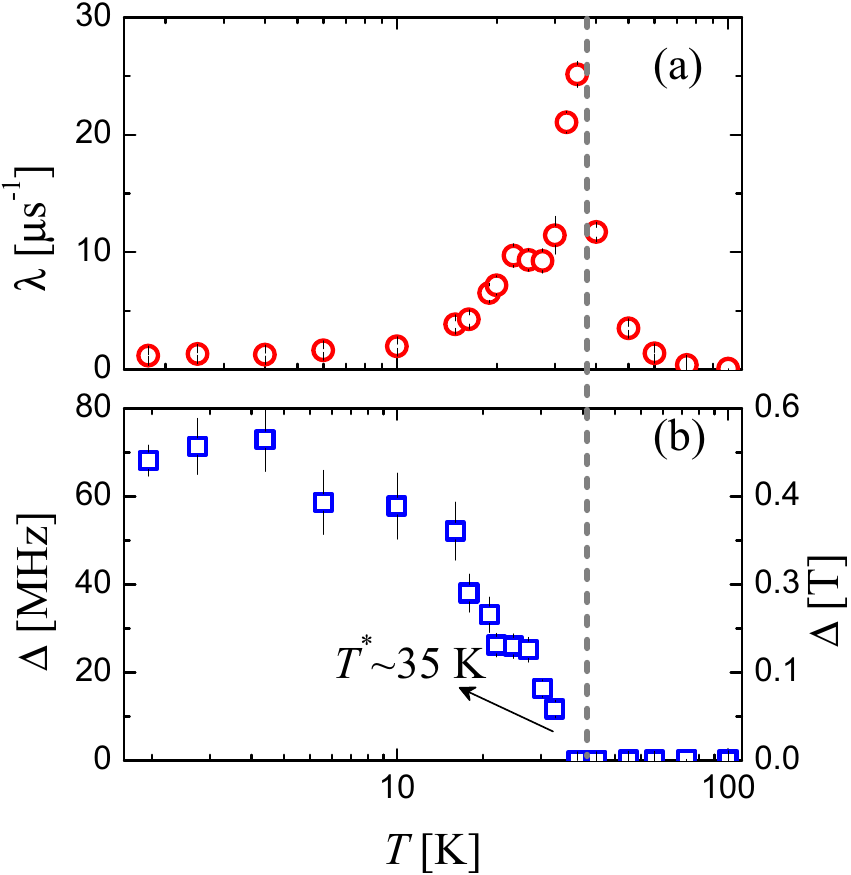}
\caption{(color online) The values of (a) $\lambda$ and (b) $\Delta$ obtained from
  fits of the relaxation curves as a function of temperature.
  \label{RlxDel}}
\end{figure}

The parameters $\lambda$ and $\Delta$ obtained from the fits are shown
in Fig.~\ref{RlxDel}(a) and (b), respectively.
Note that at high temperatures $\lambda$ is small and $\Delta \approx
0$. As the temperature is decreased $\lambda$ increases sharply while
$\Delta$ remains zero. At $T^{*}\sim 35$ K, $\lambda$ peaks and
$\Delta$ becomes non-zero. Finally, at lower temperatures $\lambda$
decreases again while $\Delta$ saturates at $\sim 70$~MHz,
corresponding to a width in field of $\sim 0.45$~T (right side axis of
Fig.~\ref{RlxDel}(b)). These temperature dependencies are commonly
seen in materials undergoing magnetic ordering at $T_c=T^*$
\cite{Kalvius01}, In such cases the increase of $\lambda$ above $T_c$
is attributed to a critical slowing down of the fluctuations and
$\Delta$ is proportional to the order parameter (local magnetization)
\cite{Kalvius01,Salman06PRB}. This behavior is quite unusual, and
exhibits a sharp transition from a paramagnetic to an ordered/static
regime, or the sudden appearance of static magnetic fields in the
system. Such static fields are rarely observed in SMMs studied by
\msr\ \cite{Branzoli09PRB}.

\begin{figure}[h]
\includegraphics[width=0.9\columnwidth]{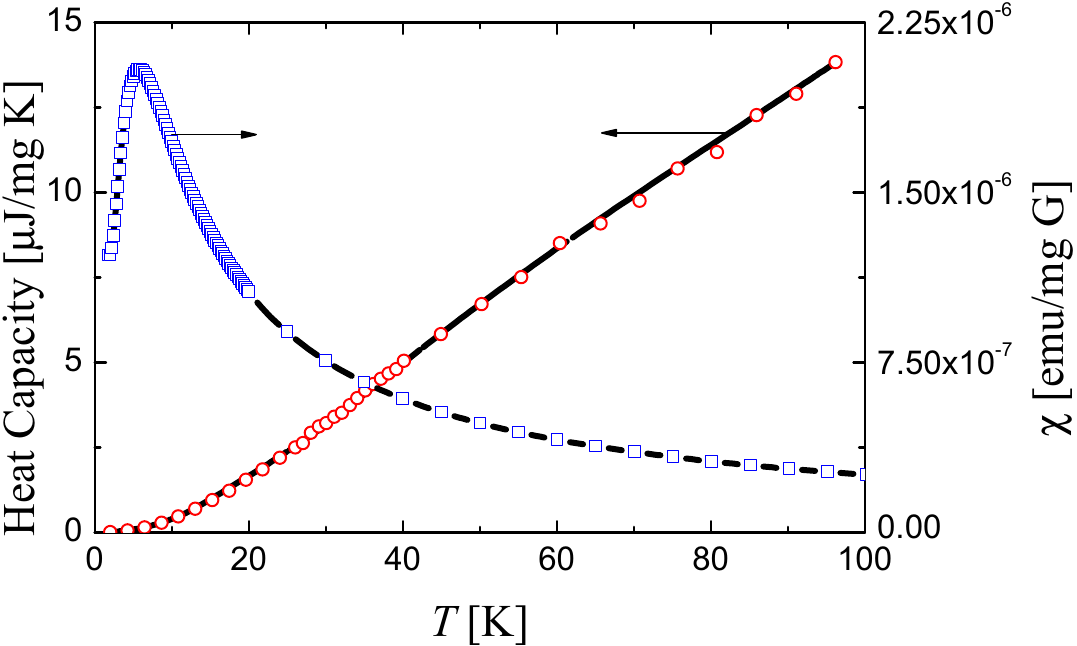}
\caption{(color online) The heat capacity measured in ZF (circles, left axes) and
  susceptibility measured in 50 mT (squares, right axes) of \Dy\ as a
  function of temperature. There is no evidence for any anomaly at 35
  K. The solid lines are a guide for the eye.
  \label{HC}}
\end{figure}

Triggered by these findings we measured the magnetization and heat
capacity to confirm that there is no evidence of long range magnetic
ordering in the \Dy\ system. As can be seen in Fig.~\ref{HC}, bulk
magnetization measurements in 50 mT show no indication of ordering at
$\sim 35$ K \cite{Tang06ACIE,Luzon08PRL}. Similarly, heat capacity
measurements in ZF show no anomalies around this temperature.
Therefore, the observed static fields can be associated with either
(I) an ordering only on the {\em local scale}, e.g. the freezing (or
slowing down) of the fluctuations of the individual \pDy\ magnetic
moments, or (II) a magnetic ground state where the \pDy\ moments order
is such a way that they do not cancel each other, resulting in a
ground state with a non-vanishing magnetic moment. The latter
possibility is ruled out by bulk magnetization measurements which
confirm a nonmagnetic ground state in \Dy\ molecules
\cite{Tang06ACIE,Luzon08PRL}.

\section{Discussion}
We start by discussing the local static field contribution and its
origin. Assuming that the muons occupy sites which are relatively far
from the \Dy\ magnetic core (i.e. at a distance much larger than the
distance between different Dy ions in the molecule), one expects that
at low temperatures the static dipolar field experienced by the muons
will decrease and vanish as the magnetic moment of \Dy\ decreases at
low temperature.  However, since muons senses a strong non-vanishing
static field at low temperatures (up to $\Delta_0 \sim 0.45$ T), we
conclude that the muons occupy sites which are close to the magnetic
core, and therefore are sensitive to the microscopic magnetic
structure of the molecules. One such possibility is a site near the Cl
ion shown in Fig.~\ref{Dy3Core}.  This site is negatively charged and
therefore may attract the positively charged implanted muons. On this
site, the dipolar fields from the 3 individual Dy ions do not cancel
each other even if the \pDy\ moments are arranged in a nonmagnetic
ground state configuration. Instead, the \pDy\ closest to the Cl
contributes mostly to the magnetic field experienced by the muon
(roughly 20 times larger). For example, assuming $\Delta_0$ is due to
a dipolar field of a single \pDy\ moment we expect that the distance
between this moment and its neighboring muon is roughly $\sim 5$ \AA.
This distance is similar to that between the Cl mentioned above and
its neighboring \pDy\ ($\sim 2.7$ \AA). Therefore, we conclude that
this static field is direct evidence of the slow relaxation of
individual \Dy\ moments, which can be associated with single \pDy\ ion
anisotropy \cite{Luzon08PRL}.

These measurements exhibits the power of a local probe measurement and
its advantages over conventional bulk magnetization measurements in
resolving the magnetic properties of individual magnetic molecules,
and the microscopic magnetic structure in general. Moreover, the
observed local static magnetic fields, despite the nonmagnetic ground
state, reflect the peculiar nature of the spin vortex arrangement and
the Ising/classical type interactions between the spins which give a
nonmagnetic doublet, i.e. two distinct magnetic states. Although muons
cannot detect the sense of chirality of the ground state, they provide
evidence of its static magnetic nature.

\begin{figure}[h]
\includegraphics[width=0.9\columnwidth]{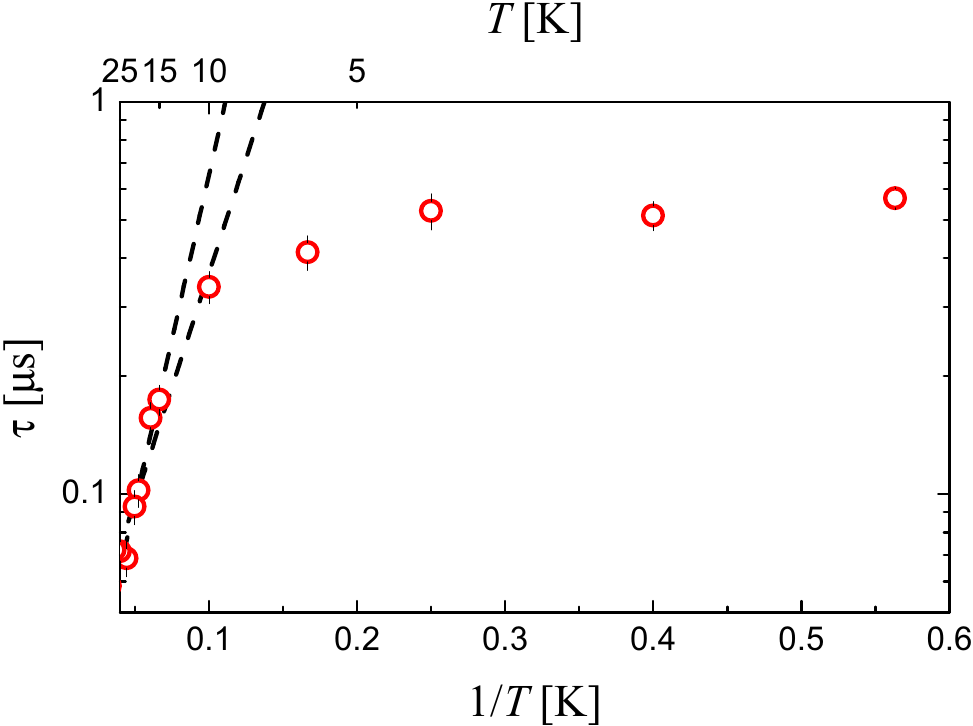}
\caption{(color online) The fluctuation time of the local field
  experienced by muons as a function of temperature. The dashed lines
  indicate the range of parameters for fits to Arrhenius
  law.\label{tauvsT}}
\end{figure}

Next, we investigate the dynamic properties of the local field
experienced by the implanted muons. As we mentioned earlier, the low
temperature relaxation $\lambda$ can be used to extract the
fluctuation time $\tau$ of the local field. In Fig.~\ref{tauvsT} we
plot $\tau$ as a function of temperatures (below $\sim 25$ K where
$\Delta>\lambda$).
We find that $\tau \sim 0.1$ $\mu$s at $\sim 20$ K, increases
gradually as the temperature is decreased and saturates at $\sim 0.55$
$\mu$s below $\sim 4$ K. These fluctuations are due to fluctuations of
the \pDy\ magnetic moments near the stopping site of the muon. In a
very simplified picture, muons are sensitive to $\tau$ only when it is
shorter than the muon's lifetime ($\tau_{\mu}=2.2$ $\mu$s). When $\tau
\gtrsim \tau_{\mu}$ we expect the muon to experience a static field
component. This is consistent with the extracted values of $\tau$; in
the range of a fraction of a $\mu$s. In contrast to this scale of
dynamics, ac susceptibility measurements of the relaxation time of the
magnetization at zero applied field have shown that it follows an
Arrhenius law \cite{Luzon08PRL}
\begin{equation} \label{Arrhenius}
\tau_m=\tau_m^0 \exp(U_m/T),
\end{equation}
where $\tau_m^0=2.5(5) \times 10^{-7}$ s and $U_m=36(2)$ K. The
similarity between the temperature dependence of our results and those
extracted from ac susceptibility is striking.  However, the low
temperature saturation value of the fluctuation time is an order of
magnitude longer in the ac susceptibility measurements.  For
comparison, fits of $\tau$ in the range 10-25 K to
Eq.~(\ref{Arrhenius}) yield an activation barrier $U=27-39$ K and
attempt time $\tau^0=(1.4-2.5) \times 10^{-8}$ s (dashed lines in
Fig.~\ref{tauvsT}). This discrepancy indicates that although both
\msr\ and ac susceptibility measurements confirm the slow relaxation
of the magnetization, the scale of dynamics measured in both
techniques is different. However, given that these techniques are very
different and have different temporal sensitivity, there are two
possible explanations of the discrepancy. One possibility is that
while the dynamics measured with ac susceptibility are those of
collective bulk magnetization, \msr\ measures fluctuations of of {\em
  individual} molecules or \pDy\ moments within it. For example,
studies in Ni$_{10}$ magnetic molecules have shown that while ac
susceptibility exhibits slow relaxation of bulk magnetization, nuclear
magnetic resonance (NMR) measurements confirm the presence of fast
single molecule dynamics \cite{Carretta06PRL,Belesi09PRL}. This was
interpreted as slowing down of the bulk magnetization due to a
resonant phonon trapping mechanism which prevents thermalization of the
bulk magnetization but allows fast spin flipping of the individual
molecular moments \cite{Belesi09PRL}. Another possibility is that each
techniques is sensitive to a different dynamic process, which may not
be even related. The dynamics of the bulk magnetization was associated
to fluctuations of the \Dy\ magnetization in the plane of the \pDy\
triangle. However, fluctuations out of the plane cannot be ruled out.
This type of fluctuations may be faster, and therefore more effective
in causing muon spin lattice relaxation. We believe that NMR
measurements in \Dy\ should be sensitive to both scales of dynamics,
and therefore would provide valuable information that could clarify
the discrepancy between bulk magnetization measurements and \msr.

\section{Conclusions}
In conclusion, we find that muons implanted in \Dy\ observe a
fluctuating field at high temperatures. Below $\sim 35$ K the local
field has two components, static and dynamic. Bulk magnetization and
heat capacity measurements rule out the existence of long range
ordering associated with the observed temperature dependence of the
local field. Instead, the observed static local magnetic field is
direct evidence of the slow relaxation of the magnetization. Our
measurements exhibit that in spite of the peculiar nature of the \pDy\
spins vortex arrangement which produces a nonmagnetic doublet, a local
probe such as muons can detect their static magnetic
nature. Therefore, although no net magnetic moment is present, they
can be used, in principle, to store magnetic information (sense of
chirality) with the advantage of long coherence time due to the small
intermolecular dipolar couplings. Finally, we find that the time scale
of low temperature dynamics detected with \msr\ is an order of
magnitude shorter than that extracted from ac susceptibility
measurements. The different time scale of dynamics could be due to the
sensitivity of the different techniques to different relaxation
processes or a difference between fluctuations of collective
individual moments.

\begin{acknowledgments}
We would like to thank Andreas Suter for reading the manuscript and
helpful discussions. This work was performed at the Swiss Muon Source
(S$\mu$S), at the Paul Scherrer Institute in Villigen, Switzerland.
\end{acknowledgments}

\newcommand{\noopsort}[1]{} \newcommand{\printfirst}[2]{#1}
  \newcommand{\singleletter}[1]{#1} \newcommand{\switchargs}[2]{#2#1}


\begin{thebibliography}{25}
\bibitem{footnote}
Simulations show that the dynamic LKT and the phenomenological function $A_{\rm LKT}(t)e^{-\lambda t}$ are identical when $\Delta > \lambda$.
\end{thebibliography}
\end{document}